\documentclass[sigconf]{acmart}

\usepackage{listings}
\usepackage{xcolor}
\usepackage{soul}
\usepackage{orcidlink}

\lstdefinestyle{cpconfig}{
  basicstyle=\footnotesize\ttfamily,
  keywordstyle=\color{blue},
  morekeywords={serviceModels, modelName, temperature, choices, chat, role, message},
  breaklines=true,
}

\AtBeginDocument{%
  }

\newcommand{\ie}{\emph{i.e., }}
\newcommand{\eg}{\emph{e.g., }}
\newcommand{\coqpilot}{\texttt{CoqPilot}}
\newcommand{\jetbrains}{JetBrains}
\newcommand{\coq}{Coq}
\newcommand{\coqlsp}{Coq-LSP}
\newcommand{\vscode}{VSCode}

\usepackage{pdfpages}
\begin{document}

\title{{\coqpilot}, a plugin for LLM-based generation of proofs}

\author{Andrei Kozyrev \orcidlink{0009-0004-3185-9368}}
\affiliation{%
  \institution{{\jetbrains} Research}
  \country{Germany}
}
\affiliation{%
  \institution{Constructor University}
  \city{Bremen}
  \country{Germany}
}

\author{Gleb Solovev \orcidlink{0009-0004-1116-7743}}
\affiliation{%
  \institution{{\jetbrains} Research}
  \country{Germany}
}
\affiliation{%
  \institution{Constructor University}
  \city{Bremen}
  \country{Germany}
}

\author{Nikita Khramov \orcidlink{0009-0004-3968-4443}}
\affiliation{%
  \institution{{\jetbrains} Research}
  \country{Germany}
}
\affiliation{%
  \institution{Constructor University}
  \city{Bremen}
  \country{Germany}
}

\author{Anton Podkopaev \orcidlink{0000-0002-9448-6587}}
\affiliation{%
  \institution{{\jetbrains} Research}
  \country{the Netherlands}
}
\affiliation{%
  \institution{Constructor University}
  \city{Bremen}
  \country{Germany}
}

\renewcommand{\shortauthors}{Andrei Kozyrev, Gleb Solovev, Nikita Khramov, and Anton Podkopaev}

\begin{abstract} 
  We present {\coqpilot}, a VS Code extension designed to help automate writing of {\coq} proofs. The plugin collects the parts of proofs marked with the \texttt{admit} tactic in a Coq file, \ie proof holes, and combines LLMs along with non-machine-learning methods to generate proof candidates for the holes. Then, {\coqpilot} checks if each proof candidate solves the given subgoal and, if successful, replaces the hole with it. The focus of {\coqpilot} is twofold. Firstly, we want to allow users to seamlessly combine multiple {\coq} generation approaches and provide a zero-setup experience for our tool. Secondly, we want to deliver a platform for LLM-based experiments on {\coq} proof generation. We developed a benchmarking system for {\coq} generation methods, available in the plugin, and conducted an experiment using it, showcasing the framework's possibilities.
  \emph{Demo of} {\coqpilot} \emph is available at: \url{https://youtu.be/oB1Lx-So9Lo}. \emph{Code at}: \url{https://github.com/JetBrains-Research/coqpilot}
\end{abstract}

\keywords{LLM, Coq, code generation}
\maketitle

\section{Introduction} \label{sec:introduction}

Testing has always been essential for making reliable software. For some specific domains, such as aerospace engineering, banking infrastructure, or medical devices, bugs in critical systems may lead to catastrophic consequences~\cite{fatalerrorsknightcapital, fatalerrorsmariner}. Formal software verification ensures that software operates correctly and safely by proving its correctness against the specification~\cite{Pierce:SF1}. Under an assumption of a well-constructed specification, formal verification provides stronger guarantees than traditional testing methods, such as unit or integration testing, due to its exhaustive nature. To date, there exist a number of interactive theorem provers (ITP), such as {\coq}~\cite{coq}, Isabelle~\cite{isabelle}, or Lean~\cite{lean}. They are designed to assist users with the construction of formal specifications and verification of formal proofs. For example, \coq helps in development by providing a robust framework for defining mathematical assertions and ensuring the logical consistency of complex formal proofs.
The formal verification approach has proved to be fruitful: for instance, CompCert~\cite{compcert}, a C compiler written in Coq, was the only C compiler in which an extensive study found no bugs~\cite{yang2011finding}.

{\coq} is an interactive proof system, where proofs are constructed step-by-step using so-called \emph{tactics}. When applied, tactics change the state of the current proof. In particular, a tactic may apply an already proven lemma, destruct the assumption to perform case analysis, apply induction reasoning, and much more. At any point of the proof, the proof state shown to the user will contain information about the current target statement and the assumptions under which it has to be proven. When the statement is empty, the proof is complete. If the proof contains an error or is not constructed correctly, {\coq}'s system will tell that the proof is invalid and provide comprehensive information on the origin of the problem.

Writing formal proofs is an exceptionally time-consuming task and requires considerable experience from the programmer~\cite{qed}. Various approaches for {\coq} generation are already present, both machine-learning-based and not. CoqHammer~\cite{coqhammer} translates the {\coq}'s logic into untyped first-order logic and searches for the proof. The K-NN approach, implemented as a back-end in Tactician~\cite{tactician}, predicts tactics based on what has been used in similar cases. Other approaches are based on generative models~\cite{tactok, coqgym, proverbot9001, rute2024graph2tac}. Recently, Large Language Models (LLMs) have gained strong code-generation capabilities~\cite{jiang2024survey}. Combined with tools for automatic code verification, we may produce high-quality, reliable code seamlessly.

Some developed models and tools for {\coq} generation may require significant setup and/or lack integration into the platform for end users~\cite{coqgym, proverbot9001, tactok}.
One other space of improvement for existing non-deterministic proof search processes is to use the information provided by the {\coq}'s system. Even for a human, writing {\coq} code in a notepad instead of a proper {\coq} IDE would be harder than in a typical programming language. Interactive stepping through each tactic invocation and updated goal states provide the necessary information while writing proofs. Fortunately, such information can be gathered automatically and used for proof generation.

We propose {\coqpilot}, a {\vscode} plugin designed to deliver a convenient generation of {\coq} code using LLMs and other methods. We studied possible external enhancements to generating {\coq} code with general-purpose models. The automatic checking of multiple generated proof candidates was developed to pick and present only the valid one to the user. We implemented premise selection for better LLM prompting and created an LLM-guided mechanism that attempts fixing failing proofs with the help of the {\coq}'s error messages. To evaluate the performance of the described solutions, we implemented a benchmarking framework for our extension. The framework allows efficiently conducting experiments on {\coq} generating using different models. We experimented with this framework, comparing several LLMs in {\coq} generation and evaluating if our contributions boost their performance. 
  
To allow automatic proof checking, we implemented a higher level module, wrapping {\coq} Language Server%
\footnote{Language Server Protocol: \url{https://microsoft.github.io/language-server-protocol/}}
and providing abstractions such as the one to check if the given proof for the theorem is valid in a particular environment. We used the particular {\coq} language server implementation~\cite{emiliocoqlsp}, and from now onwards, we will refer to it as {\coqlsp}.

The proposed {\coqpilot}'s architecture is modular regarding the target language, requiring minimal code changes to adapt to another language. {\coqpilot} integrates popular LLMs and allows users to include tools like Tactician and CoqHammer in the proof generation pipeline. During implementation, we addressed the challenges of using commercial LLMs, including managing token limits and handling failures, by developing mechanisms for retries and clear user feedback.

In Section \ref{sec:coqpilot}, we describe the plugin and the challenges we overcame during its development process. Section \ref{sec:benchmark} discusses the created benchmarking framework. In Section \ref{sec:evaluation}, we describe the experiment we conducted and evaluate the features proposed in Section \ref{sec:coqpilot}. In Section \ref{sec:relatedwork}, we cover related work, and in Section \ref{sec:conclusion}, we conclude and glance at future work. 

\section{CoqPilot} \label{sec:coqpilot}

In {\coq}, a goal represents a statement or proposition to be proven. One typically starts the proof with the statement you want to establish as true. Then, one applies tactics and transforms the current goal into one or more subgoals that are simpler or more manageable. Special \texttt{admit} tactic allows skipping a subgoal to permit further progress on the rest of the proof. If the proof contains admits, it is considered incomplete. Each \texttt{admit} corresponds to a self-contained goal with the hypotheses and the conclusion. Say we have a {\coq} file with a number of unfinished proofs containing admits. {\coqpilot} runs over the admits and tries to substitute them with correct proofs.

We designed {\coqpilot} to serve as a tool for combining different approaches to {\coq} generation. We implemented multiple ways to fetch completion and infrastructure around {\coqlsp} to check proof candidates. We will refer to each way of fetching completion as a \emph{service}. Currently, the available services are OpenAI API, LLMs running locally through LMStudio, {\jetbrains} AI Platform, and completion via predefined automation tactics. {\coq} automation tools such as Tactician~\cite{tactician} and CoqHammer~\cite{coqhammer}, which are triggered using special tactics, could be added to the pipeline through the predefined tactics to unite generation capabilities with {\coqpilot}. 

A particular setup for the completion request is denoted as \emph{model parameters}. For LLM-based services, model parameters include the LLM name, the temperature, the prompt, the number of choices to make (\ie the number of completions the model should generate), and other specifications. As LLMs are not deterministic, making several attempts for each model is beneficial. This result is backed up in Section \ref{sec:evaluation}. The setup of {\coqpilot} consists of a list of model parameters for each of the chosen services. 

While implementing the described approach, we encountered several difficulties that affected the {\coqpilot}'s final architecture. Different proof holes in {\coq} have independent states, and we intend to generate completion for distinct holes in parallel. This requires introducing safety of concurrency to our developed proof-checking mechanisms since {\coqlsp} cannot process parallel requests. Our goal was to develop an infrastructure with interchangeable components to allow users to easily add new services and prepare the ground to interchange {\coq} with another ITP.

Accurate error handling presents other challenges. Services such as OpenAI have various types of errors, which are supposed to be handled differently. Some may be classified as parameter validation errors and presented to the users; others may be service errors. One of the critical failures, which mainly occurs during benchmarking, is caused by exceeding token limits. Commercial LLM providers restrict their models' usage rates. Local token counters are imprecise, which makes it challenging to overcome these limitations. Therefore, correctly handling such errors becomes crucial for presenting them to the user in an understandable manner. To address this issue, we developed a custom error class hierarchy, differentiating between configuration errors, generation failures, and connection errors, and repacked specific service errors into these appropriate classes. The implementation reports and logs errors based on their types, supporting both user and benchmarking modes. 

{\coqpilot} offers many configurable parameters. They help the user to set up both the plugin behavior and the experiments using the benchmark. We have implemented a parameter resolution framework, which correctly handles errors, allowing a programmer to write reliable resolvers for new services and parameters.

One of our contributions is enhancing the capabilities of general-purpose LLMs in generating {\coq} code. Given a position to perform completion, we can get the desired statement to prove and the hypotheses under which it should hold. However, this is usually not enough. Writing {\coq} proofs, a human often recalls other lemmas and objects in the corresponding file/project.
It may be challenging for the model to deal with a theorem isolated from its context. 
To address this problem, we perform premise selection%
\footnote{Retrieval of facts from some given knowledge base that can help the model and advance the proof.}
for the theorems within the same file and use them as a few-shot prompt to an LLM. During few-shot prompting, several concrete examples of how the task needs to be solved are provided. 

Few-shot prompting gives the model a better understanding of the problem context and structures the format of its output. Due to token limitations and the model's context window size, we can usually only take a subset of theorems from the file. We choose optimal premises using metrics such as distance from the generation target or similarity with other theorem statements.

Also, we may extract helpful information from the {\coq}'s system when proof candidates fail. In particular, we may get the error that occurred and use it to try to fix the failing proof. Baldur~\cite{first2023baldur} used an idea of proof repairing to train a separate proof fixing model. A similar to {\coqpilot} approach with general purpose LLMs may be found in Copra~\cite{copra}. When {\coqpilot}'s general pipeline does not find the proof, we launch a multi-turn communication process with an LLM. The number of completions to fetch per turn and the depth $d$ are predefined in settings by the user. We send the compilation error and special prompt to the LLM and ask it to fix it. If the proof is still not accepted by {\coq} afterward, we repeat the process, but at a maximum of $d - 1$ times.  

\section{CoqPilot benchmark} \label{sec:benchmark}

We aimed to develop a benchmarking framework to evaluate the current effectiveness of features implemented in {\coqpilot} and find space for further improvements. Specifically, our research questions included (i) how well general purpose LLMs can write {\coq} proofs, (ii) to which extent does {\coqpilot} improve the LLM approach to {\coq} generation, and (iii) which additional value \coqpilot, using general-purpose LLMs, contributes to other {\coq} automation tools such as CoqHammer and Tactician?

Implementing such a framework brought up several issues that we solved. The main peculiarity of our benchmarking approach is that we need to send a large number of tokens to each model. In order to maintain reasonable performance, we aim to make requests as fast as possible. However, there is usually a limitation on the number of tokens that can be sent in a short time frame. To overcome this, we considered the requests to each model in each service to complete the given goal as a separate asynchronous \emph{task}. We heuristically determined the necessary \emph{waiting} time for these tasks to comply with the mentioned limitation. 

The developed benchmark framework provides a number of possibilities. First, it allows the gathering of information about the internal state of {\coqpilot}, {\eg} 
the theorems chosen for the context and the number of used tokens.  Second, thanks to the implemented interfaces and the {\coqpilot's} architecture, the developed framework can be conveniently scaled for experimenting with other tools. In this work, we experimented with Tactician and CoqHammer. Moreover, it is possible to generate tailored reports based on the results of the experiments that were conducted.

\section{Evaluation} \label{sec:evaluation}

To evaluate the performance of {\coqpilot}, we required a dataset with a large number of human-written theorems and proofs. As it was said before, {\coqpilot} depends on {\coqlsp}, which is not version agnostic and supports {\coq} versions starting from 8.15. Due to this limitation, it was impossible to fully leverage the CoqGym dataset~\cite{coqgym} for our experiments as it contains projects requiring older {\coq} versions. We have decided to limit ourselves to {\coq} 8.19 as the latest version available. We have chosen the IMM project\footnote{IMM: \url{https://github.com/weakmemory/imm}} for our experiment. The project consists of a large number of proofs and supports {\coq} 8.19. Moreover, the IMM is of particular interest to our lab since it is developed there.

The data for the experiments was prepared as follows. We decided to consider only the proofs of at most 20 tactics, as we initially developed {\coqpilot} to help users generate subgoals or smaller lemmas. Theorems with proofs with such lengths amount to 83\% of proofs in the IMM project. Due to the amount of computing and financial resources at our disposal, we have been unable to experiment on the entire project. Therefore, we decided to use a relatively small subset of 300 theorems.
Moreover, we wanted to split the dataset into three groups based on the length of proofs measured in tactics. This was done to provide a clearer interpretation of the results. The group sizes were chosen with respect to the distribution of proof lengths in the given project so that the results of the experiments would be representative of the entire project. The final group sizes are presented in Table \ref{tab:freq}.

During the main experiment, we evaluated how many theorems from the constructed dataset could be proven using different methods. We used {\coq}'s built-in first-order reasoning tactic with automation \texttt{firstorder auto with *} as a baseline. We have chosen GPT-4o, GPT-3.5, Anthropic Claude, and the open-sourced LLaMA-2 13B Chat as models. The average number of theorems sent to a model varied from the context window size, \eg for GPT-4o it was 52. Completion choices were equal to 12 for GPT-4o, 20 for GPT-3.5 and LLaMA and 7 for Anthropic Claude. The multi-round feature was disabled due to the exhaustive tokens consumption of this feature. Along with the models listed above, we have tested Tactician and CoqHammer with timeouts of 30, 60, and 90 seconds for the three groups, respectively. If the proof was not found during the specified timeout, we consider the theorem as unsolved. The percentages in the table cells represent the proportion of theorems in each group successfully solved using the specified method.
More details are provided in the experiment report.%
\footnote{\url{https://github.com/JetBrains-Research/coqpilot/tree/main/etc/docs/benchmark}} 

\begin{table}[htp]
  \caption{Benchmarking results}
  \label{tab:freq}
  \begin{tabular}{lrrrr}
    \toprule
    Reference proof length & $\leqslant 4$ & 5--8 & 9--20 & Total \\
    Group size   & 131& 98& 71& 300 \\
    \midrule
    \texttt{firstorder auto with *} & 11\% & 2\% & 1\% & 6\% \\
    \midrule
    OpenAI GPT-3.5          & 29\% & 17\% & 6\% & 20\% \\
    OpenAI GPT-4o           & 50\% & 26\% & 15\% & 34\% \\
    LLaMA-2 13B Chat        & 2\% & 0\% & 0\% & 0.5\% \\
    Anthropic Claude        & 21\% & 7\% & 7\% & 13\% \\
    All models together     & 57\% & 32\% & 18\% & 39\% \\
    \midrule
    Tactician               & 45\% & 23\% & 10\% & 29\% \\
    CoqHammer               & 23\% & 4\% & 0\% & 11\% \\
    \midrule
    All methods together    & 71\% & 45\% & 23\% & 51\% \\
    \bottomrule
\end{tabular}
\end{table}

\begin{figure}[htp]
    \caption{Venn diagram of the proven theorems}
    \label{fig:venn}
    \centering
    \includegraphics[height=4.5cm]{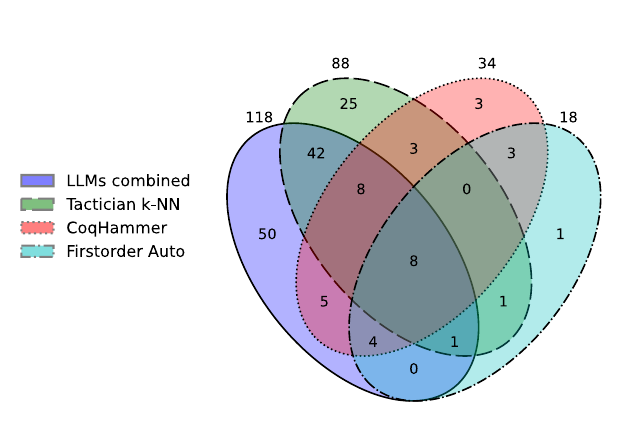}
\end{figure}

GPT-4o with {\coqpilot}'s approach can prove 34\% theorems, as seen in Table \ref{tab:freq}, with 51\% of them being proved on the first attempt. GPT-3.5 can only prove 30\% on the first attempt, which can be explained by the smaller context window size, resulting in fewer chosen premises. Another noticeable result is that among each group, the collectible effort of all models is stronger than any individual one. It shows that the approach of {\coqpilot} to using a sequence of different models altogether is promising. A combination of four models used through {\coqpilot}, CoqHammer, and Tactician can prove 51\% theorems. The user can invoke this powerful combination from {\coqpilot} with a single call. Figure \ref{fig:venn} shows methods corresponding to the sets of theorems they can prove. 

Additionally, to examine how varying the number of premises sent to the model impacts the results, we compared GPT-4o with a different number of premises used in context on another 50 samples from the IMM project. Results show that the model can solve 0\% with 0 theorems as premises, 8\% with 1 theorem, and 32\% theorems with the maximum possible number of premises.%
\footnote{The maximum possible number of premises is calculated as the maximal number of premises that fit into the model's context window.}
Another experiment with GPT-4o and 50 theorems showed that the multi-round mechanism with depth 2 and width 2 fixes 2 proofs in addition to the ones that were already generated correctly. 

The results demonstrate the benefits of using {\coqpilot} instead
of plain LLMs, showcase additional value to other {\coq} provers, and highlight the usability of using multiple generation methods at once via {\coqpilot}.

\section{Related work} \label{sec:relatedwork}

Many {\coq} generation tools require a long setup and are hardly integrated into the {\coq} development workflow. Proofster~\cite{proofster} is a web interface for {\coq} proof synthesis and exploration. Tactician~\cite{tactician} tries to generate {\coq} proofs after invocation by the special tactics. Copra~\cite{copra} is an agent for theorem proving, which repeatedly uses a general-purpose LLM for completion. Our work serves similar purposes with a focus on a couple of factors. We aim to develop a plugin that incorporates well into a typical user's workflow and provides setup-free experience. We have built our tool around uniting many approaches and seamlessly allowing users to try all available tools for their problems. This pipeline also brings convenience in experimenting. Another focus is automatically boosting non-deterministic {\coq} generation tools with the {\coq}'s proof checker. Along with that, we implemented fetching completion from common LLM providers. Tools such as Tactician can be used in {\coqpilot} as services via predefined tactics without any additional effort from the user.

\section{Conclusion} \label{sec:conclusion}

We presented {\coqpilot}, a {\vscode} plugin for {\coq} generation that requires minimal setup. We allow users to seamlessly switch between different {\coq} generation methods and easily add new ones. We contributed techniques that boost the performance of general-purpose LLMs. Compared to one-shot plain GPT-4o invocation, which can solve 0\% theorems from the compiled dataset, GPT-4o with {\coqpilot}' modifications gets 34\%. As shown in Table \ref{tab:freq}, the joint effort of four models integrated into ${\coqpilot}$, achieves 39\%. We contributed a highly configurable experiment framework for 
testing methods implemented in {\coqpilot} for {\coq} generation. 

\begin{acks}
This paper has been greatly improved by the comments of Yaroslav Golubev, Ekaterina Verbitskaia and Marat Akhin. We thank Emilio Jesús Gallego Arias for his work on {\coqlsp}. 
\end{acks}

\bibliographystyle{ACM-Reference-Format}
\bibliography{CoqPilot24_PrePrint_ArXiv}

\end{document}